\documentclass[conference]{IEEEtran}
\IEEEoverridecommandlockouts

\usepackage{cite}
\usepackage{amsmath,amssymb,amsfonts}
\usepackage{graphicx}

\usepackage{tabularx}
\usepackage{amssymb}
\usepackage{pifont}
\usepackage{array}
\usepackage{multirow}
\usepackage{algorithm}
\usepackage{algorithmicx}
\usepackage{algpseudocode}

\def\BibTeX{{\rm B\kern-.05em{\sc i\kern-.025em b}\kern-.08em
    T\kern-.1667em\lower.7ex\hbox{E}\kern-.125emX}}
\begin{document}

\title{PseudoVC: Improving One-shot Voice Conversion with Pseudo Paired Data}

\author{
\IEEEauthorblockN{
{\begin{tabular}{c}
        \textit{Songjun Cao, Qinghua Wu, Jie Chen, Jin Li, Long Ma}
    \end{tabular}}}
\IEEEauthorblockA{Tencent Youtu Lab, China}
}

\maketitle

\begin{abstract}


As parallel training data is scarce for one-shot voice conversion (VC) tasks, waveform reconstruction is typically performed by various VC systems. A typical one-shot VC system comprises a content encoder and a speaker encoder. However, two types of mismatches arise: one for the inputs to the content encoder during training and inference, and another for the inputs to the speaker encoder.
To address these mismatches, we propose a novel VC training method called \textit{PseudoVC} in this paper. First, we introduce an innovative information perturbation approach named \textit{Pseudo Conversion} to tackle the first mismatch problem. This approach leverages pretrained VC models to convert the source utterance into a perturbed utterance, which is fed into the content encoder during training. Second, we propose an approach termed \textit{Speaker Sampling} to resolve the second mismatch problem, which will substitute the input to the speaker encoder by another utterance from the same speaker during training.
Experimental results demonstrate that our proposed \textit{Pseudo Conversion} outperforms previous information perturbation methods, and the overall \textit{PseudoVC} method surpasses publicly available VC models. Audio examples are available \footnote{https://songjuncao.github.io/pseudovc/}.

\end{abstract}

\section{Introduction}
\label{sec:introduction}
Voice conversion (VC) is a task that transfers the voice of a source speaker to a target speaker, while maintaining the linguistic information. One-shot voice conversion is a special case where only one utterance of the target speaker is available. Recently, many works \cite{choi2024dddm,li2023freevc,shan2024phoneme,baas2023voice,casanova2022yourtts,wang2024gr0} has been proposed to advance the development of the one-shot VC task.

\begin{figure}[t]
  \centering
  \includegraphics[width=\linewidth]{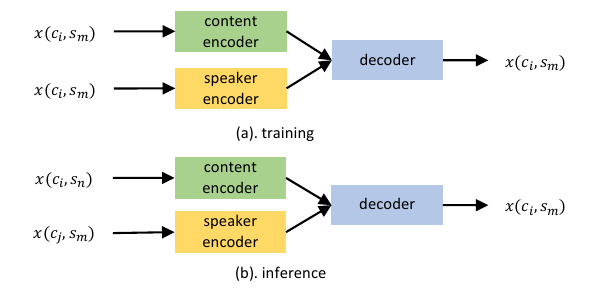}
  \caption{Mismatch between training and inference}
  \label{fig:mismatch}
\end{figure}

As parallel training data is hard to achieve \cite{kaneko2017parallel,tian2018average,lorenzo2018voice}, many VC models are trained by reconstructing the source speech.
Those systems \cite{choi2024dddm,li2023freevc,wang2024gr0} typically consist of a content encoder, a speaker encoder and a decoder \footnote{Here we simply the model structure, as other information (pitch) can be modeled independently in some works.}, as depicted in Fig.~\ref{fig:mismatch}.
In order to illustrate the mismatch problem more intuitively, we assume the outputs of both training and inference to be $x(c_i, s_m)$, where $c_i$ and $s_m$ denote content information and speaker information, respectively.
The content encoder extracts linguistic information from the source utterance $x(c_i, s_m)$, where $c_i$ and $s_m$ denote content information and speaker information, respectively.
The speaker encoder extracts speaker-related information from the source utterance $x(c_i, s_m)$.
The decoder then reconstructs the source utterance $x(c_i, s_m)$ based on the extracted content and speaker information.
However, there are two mismatches between training and inference phases: 1. The inputs to the content encoder during training and inference are $x(c_i, s_m)$ and $x(c_i, s_n)$. 2. The inputs of the speaker encoder during training and inference are $x(c_i, s_m)$ and $x(c_j, s_m)$.

To alleviate the first mismatch problem, some information perturbation approaches \cite{li2023freevc,qian2020unsupervised,chan2022speechsplit2,choi2021neural} are proposed to transform the input of the content encoder from $x(c_i, s_m)$ to $x'(c_i, s_n)$ during training.
This transformation aims to modify the speaker-related information while preserving the content information $c_i$.
Although these perturbation methods can mitigate the mismatch problem to some extent, they still exhibit limitations in the diversity and naturalness of the generated $x'(c_i, s_n)$, which can adversely affect the performance of voice conversion.
To overcome this problem, we propose an effective method termed \textit{Pseudo Conversion} in this paper.
This method utilizes pretrained VC models to convert the source utterance $x(c_i, s_m)$ into a pseudo utterance $x'(c_i, s_n)$, which will be fed into the content encoder.
The timbre of generated pseudo utterances $x'(c_i, s_n)$ can be diverse and natural, thereby reducing the gap between training and inference.

Regarding the second mismatch problem, there has been relatively little exploration in the literature \cite{choi2024dddm}.
Inspired by scheduled sampling \cite{bengio2015scheduled}, we propose a simple method called \textit{Speaker Sampling}.
During training, we replace the input of the speaker encoder from $x(c_i, s_m)$ to  another utterance spoken by the same speaker with a certain probability.

\section{Related Work}
\label{sec:related}
For the perturbation on the input of content encoder, SPEECHSPLIT2.0 \cite{chan2022speechsplit2} uses Vocal Tract Length Perturbation (VTLP) \cite{jaitly2013vocal} to modify the timbre by warping the frequency.
NANSY \cite{choi2021neural} introduces a method that perturbs the information of the source utterance through three signal processing techniques.
Additionally, \cite{li2023freevc} utilizes an SR-based data augmentation approach to distort speaker information in the source utterance, thereby assisting the model in learning to extract clean content information.
All these methods aims to modify the timbre of the source waveform, while preserving the linguistic information.
However, despite these efforts, there remains a gap in terms of naturalness and speaker diversity between the perturbed utterances and real utterances.

DDDM-VC \cite{choi2024dddm} mixes the speaker representation by using a binary selection between the original and shuffled representations in the same batch. So the speaker of selected representation may be different from the source utterance.
In contrast, our approach for the speaker encoder input involves using other utterances from the same speaker during training. 

\section{Proposed Method}
\subsection{Overall architecture}
\label{sec:overall}

\begin{figure}[h]
  \centering
  \includegraphics[width=\linewidth]{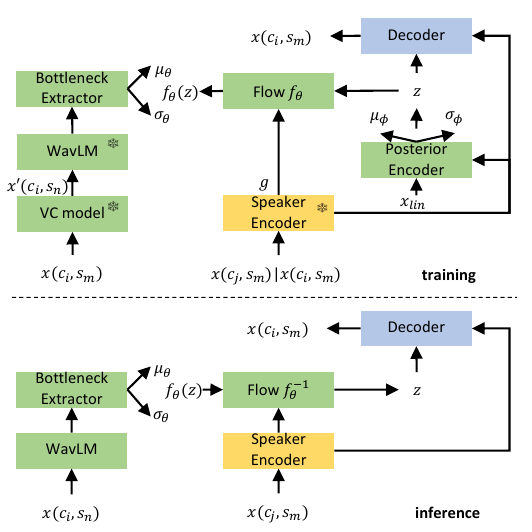}
  \caption{Top: The model architecture during training. Bottom: The model architecture during inference.}
  \label{fig:overall}
\end{figure}

As illustrated in Fig.~\ref{fig:overall}, our model is based on FreeVC \cite{li2023freevc}.
The content encoder of our model contains a WavLM model \cite{chen2022wavlm}, a bottleneck extractor and a normalizing flow \cite{rezende2015variational}.
For the speaker encoder, we utilize a pretrained speaker verification model that has been trained on a large-scale corpus \cite{liu2021any}.
The posterior encoder is composed of non-causal WaveNet residual blocks used in WaveGlow \cite{prenger2019waveglow}.
The decoder is the HiFIi-GAN V1 generator \cite{kong2020hifi}.
%
\begin{gather}
  L_{total}=L_{rec}+L_{kl}+L_{adv}(D)+L_{adv}(G)+L_{fm}(G)
    \label{equation:total}
\end{gather}

Following \cite{li2023freevc}, the training loss is defined by Equation \ref{equation:total}.
This loss function comprises several components: the KL divergence loss $L_{kl}$, the reconstruction loss $L_{rec}$, the adversarial loss $L_{adv}(D)$ for the discriminator $D$, the adversarial loss $L_{adv}(G)$ for the generator $G$ \cite{mao2017least}, and the feature-matching loss $L_{fm}(G)$ \cite{larsen2016autoencoding}.

\subsection{Training strategy}

\subsubsection{Pseudo Conversion}
Inspired by semi-supervised learning \cite{sohn2020fixmatch, zhang2022censer, arazo2020pseudo}, we propose a novel information perturbation method to alleviate the first mismatch problem defined in Section \ref{sec:introduction}.
Firstly, we train a one-shot VC model $\mathcal{M}_t$ following \cite{li2023freevc}, which is named as the VC teacher model.
Then, we generate pseudo utterances using $\mathcal{M}_t$ through the following process.
\begin{gather}
  x'(c_i, s_n) = \mathcal{M}_t(x(c_i, s_m), x(c_j, s_n))
    \label{equation:convert} \\
  \mathcal{S}(x(c_i, s_m))=\{x'(c_i, s_n)\}, n=1,...,N \label{equation:set}
\end{gather}
Assuming the source utterance is $x(c_i, s_m)$, where $c_i$ and $s_m$ denotes the content and speaker.
We can convert the source utterance to a pseudo utterance $x'(c_i, s_n)$ using Equation \ref{equation:convert}.
Here, $x(c_j, s_n)$ denotes a reference utterance with speaker $s_n$, which is randomly selected from the training dataset.
$\mathcal{S}(x(c_i, s_m))$ is the set of pseudo utterances corresponding to the source utterance $x(c_i, s_m)$, and $N$ indicates the total number of generated pseudo utterances.

The constructed $(x'(c_i, s_n), x(c_i, s_m))$ are referred to as \textit{pseudo paired data}.
During training, one pseudo utterance $x'(c_i, s_n)$ 
 will be randomly selected from $\mathcal{S}(x(c_i, s_m))$ and fed into the WavLM model, while the source utterance $x(c_i, s_m)$ serves as the target for prediction.

The above process of generating pseudo utterances is termed \textit{Pseudo Conversion}.
While it draws inspiration from traditional semi-supervised learning, there are notable differences in the details.
In the semi-supervised learning, pseudo labels generated by the teacher model are used as the model's outputs.
In our method, pseudo utterances generated by the teacher model are used as the model's inputs.
This approach aims to enhance the robustness of the model by providing it with diverse pseudo paired data during training.

\subsubsection{Speaker Sampling}
As shown in Fig.~\ref{fig:overall}, the speaker embedding $g$ extracted by the speaker encoder plays a vital role in our model.
It will be fed into the Flow, Decoder and Posterior Encoder components.

However the second mismatch problem defined in Section \ref{sec:introduction} will affect the performance.
Similar mismatch problem exists in sequence prediction task \cite{bengio2015scheduled}, which can be alleviated by scheduled sampling.
Similarly, we propose a simple training strategy to alleviate this problem, which is called \textit{Speaker Sampling}.
In this approach, we modify the input to the speaker encoder from $x(c_i, s_m)$ to $x(c_j, s_m)$ during training with a certain probability $\alpha$. 
Here, $x(c_j, s_m)$ is randomly selected from other utterances of the same speaker $s_m$.

The overall training process combining \textit{Pseudo Conversion} and \textit{Speaker Sampling} is formulated in the Algorithm \ref{alg:strategy}.

\begin{algorithm}[h]
\caption{Training Process of PseudoVC} \label{alg:strategy}
\begin{algorithmic}[1] 
\State \textbf{Input:} Training utterance $x(c_i, s_m) \in \mathcal{X}$. Generated pseudo utterance number $N$. Training iteration number $Q$. The probability of speaker sampling $\alpha$.
\State \textbf{Output:} Trained VC model $\mathcal{M}$
\State Train $\mathcal{M}_t$ on $\mathcal{X}$ following \cite{li2023freevc}
\For{each utterance $x(c_i, s_m)$ in $\mathcal{X}$}
    \For{$n=1$ to $N$}
        \State Randomly select utterance $x(c_j, s_n)$ from $\mathcal{X}$
        \State Generate pseudo utterance $x'(c_i, s_n)$ via Eq. \ref{equation:convert} \label{alg:pseudo}
    \EndFor
    \State Combine pseudo utterances into set $\mathcal{S}(x(c_i, s_m))$
\EndFor

\For{$q = 1$ to $Q$}
    \State Randomly draw a batch $\mathcal{B}$ from $\mathcal{X}$
    \For{each utterance $x(c_i, s_m)$ in $\mathcal{B}$}
        \State Randomly select $x'(c_i, s_n)$ from $\mathcal{S}(x(c_i, s_m))$
        \State $x'(c_i, s_n)$ will be fed into the content encoder
        \State Generate a random number $r$ in the range $[0, 1]$
        \If{$r < \alpha$}
            \State $x(c_j, s_m)$ will be fed into the speaker encoder
        \Else
            \State $x(c_i, s_m)$ will be fed into the speaker encoder
        \EndIf
    \EndFor
    \State Train VC model $\mathcal{M}$ with loss of Equation \ref{equation:total}
\EndFor
\State \Return trained model $\mathcal{M}$
\end{algorithmic}
\end{algorithm}
%

\section{Experiment}
\subsection{Experimental Setup}
Our experiments are conducted on VCTK \cite{Veaux2017CSTRVC} and LibriTTS \cite{zen2019libritts}.
We use VCTK corpus for training.
The split of training validation and testing follows \cite{li2023freevc} exactly.
The test-clean subset of LibriTTS is used for test.
We resample the audio samples from 24 kHz to 16 kHz.

We use the pre-trained WavLM-large encoder to extract 1024-dimensional vectors for every 20ms of 16 kHz audio.
We adopt the pretrained speaker model of \cite{liu2021any} as speaker encoder, which is fixed during training. 
Our model is trained up to 200k steps using 8 NVIDIA V100 GPUs.
The batch size is set to 64, with a maximum segment length of 128 frames.
We set $N$ of Equation \ref{equation:set} to 25,  meaning that 25 pseudo utterances will be generated for each source utterance.

\subsection{Baseline models}
For the overall performance comparison, we choose three public models, which are: 
\begin{enumerate}
\item \textbf{PH} \cite{shan2024phoneme}, a one-shot VC model named Phoneme Hallucinator. It is based on set expansion. The number of hallucinated features is set to 30000.
\item \textbf{DDDM-VC} \cite{choi2024dddm}, a recently proposed diffusion-based generative model.
\item \textbf{FreeVC} \cite{li2023freevc}, a widely-used voice conversion framework.
It is the exact baseline of our PseudoVC, as the two methods adopt similar model framework and training data. 
\end{enumerate}

For the ablation study of our proposed \textit{Pseudo Conversion} method, we select three information perturbation methods for comparison. Each of those methods generates 25 perturbed utterances:
\begin{enumerate}
\item \textbf{VTLP} is a signal processing method used by \cite{jaitly2013vocal}. The warping factor is set to $\alpha \sim U(0.9,1.1)$.
We use the implementation \footnote{https://github.com/biggytruck/SpeechSplit2.git}.
\item \textbf{NANSY} is proposed in \cite{choi2021neural}. The method has three functions, which are random frequency shaping, pitch randomization and formant shifting. The hyperparameters of those functions are consistent with those in \cite{choi2021neural}.
We refer to the project \footnote{https://github.com/dhchoi99/NANSY.git}.
\item \textbf{SR} is used by \cite{li2023freevc}. This conversion is based on vertical spectrogram-resize operation. As described by \cite{li2023freevc}, the resize ratio ranges from 0.85 to 1.15.
The details can be found in the official FreeVC project \footnote{https://github.com/OlaWod/FreeVC.git}.
\end{enumerate}

\subsection{Evaluation Metrics}
\subsubsection{Objective metrics}
We utilize Amphion \cite{zhang2023amphion} to calculate the word error rate (WER) using the Whisper large-v3 model \cite{radford2023robust}.
Additionally, we measure speaker similarity through speaker encoder cosine similarity (SECS) using the Resemblyzer \cite{wan2018generalized} model.
Following \cite{li2023freevc}, we randomly select 400 utterances (200 from VCTK, 200 from LibriTTS) as source speech, and we choose 12 speakers as target speakers, which results in 4800 utterances.
\subsubsection{Subjective metrics}
For the subjective metrics, we measure the 5-scale mean opinion score (MOS) and similarity mean opinion score (SMOS) for the speech naturalness and speaker similarity. 1 is the lowest perceived quality and 5 is the highest perceived quality.
15 participants evaluate the scores of 300 utterances which are randomly selected from the objective test dataset.

\subsection{Main results}
\begin{table}[t]
\caption{Subjective evaluation results in terms of 5-scale MOS and SMOS with 95\% confidence intervals. Objective evaluation results in terms of  WER and SECS. GT denotes the source speech.}
\begin{center}
\begin{tabular}{|c|c|c|c|c|}
\hline
Models & MOS($\uparrow$) & SMOS($\uparrow$) & WER($\downarrow$) & SECS($\uparrow$) \\
\hline
GT & 4.45 $\pm$ 0.10 & - & 3.3 & - \\
\hline
PH & 3.92 $\pm$ 0.10 & 3.44 $\pm$ 0.11 & 7.8 & 0.661 \\
\hline
DDDM-VC & 3.78 $\pm$ 0.12 & 3.63 $\pm$ 0.11 &12.6 & 0.735 \\
\hline
FreeVC & 4.25 $\pm$ 0.08 & 3.88 $\pm$ 0.10 & 6.7 & 0.751 \\
\hline
PseudoVC & \textbf{4.35 $\pm$ 0.08} & \textbf{4.06 $\pm$ 0.09} & \textbf{6.1} & \textbf{0.778} \\
\hline
\end{tabular}
\label{tab:main}
\end{center}
\end{table}

The subjective and objective results in Table \ref{tab:main} show that our proposed PseudoVC model outperforms all the baseline models.
When compared to the exact baseline FreeVC, our proposed model demonstrates superior naturalness and speaker similarity across both subjective and objective metrics.
We attribute these improvements to our two effective training strategies, which aims to reduce the mismatch between training and inference.

In the next section, we will conduct an ablation study to further investigate the effectiveness of the two training strategies.

\subsection{Ablation study of Pseudo Conversion}

\begin{table}[t]
\caption{Ablation study of Pseudo Conversion. \textbf{Information Perturbation} stands for information perturbation methods. $\mathcal{M}_t$ denotes the VC teacher model which is used by \textit{Pseudo Conversion} to generate pseudo utterances.}
\begin{center}
\begin{tabular}{|c|c|c|c|c|}
\hline
{Models} & {Information Perturbation} & $\mathcal{M}_t$ & {WER($\downarrow$)} & {SECS($\uparrow$)} \\
\hline
a0 & - & - & 7.1 & 0.736 \\
\hline
a1 & VTLP & - & 6.5 & 0.758 \\
\hline
a2 & NANSY & - & 6.7 & 0.765 \\
\hline
a3 & SR & - & 6.7 & 0.751 \\
\hline
b1 & Pseudo Conversion & a1 & 6.4 & 0.776 \\
\hline
b2 & Pseudo Conversion & a2 & 6.5 & 0.778 \\
\hline
b3 & Pseudo Conversion & a3 & 6.7 & 0.780 \\
\hline
\end{tabular}
\label{tab:ablation1}
\end{center}
\end{table}

\subsubsection{Comparison with previous work}
In Table \ref{tab:ablation1}, we compare our proposed \textit{Pseudo Conversion} with other information perturbation methods.
All the models of Table \ref{tab:ablation1} only differ in the input of WavLM.
\textbf{a0} is the baseline model, which feeds source utterances into the WavLM without any perturbation.
Results show that all the information perturbation methods achieve better results than the baseline model \textbf{a0}.
Model \textbf{b3} using \textit{Pseudo Conversion} achieves the best speaker similarity among those methods.

\begin{figure}[t]
  \centering
  \includegraphics[width=0.5\textwidth]{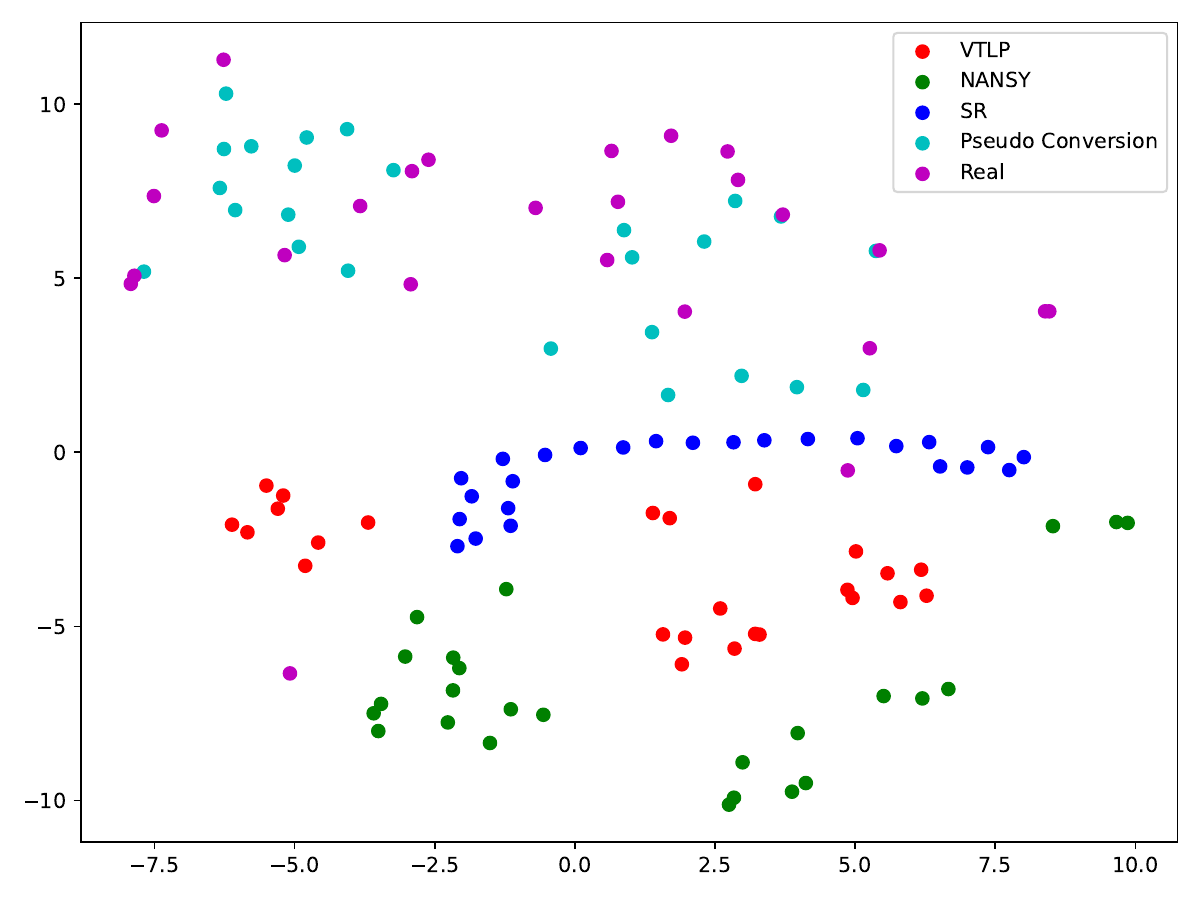}
  \caption{Visualizations of speaker embeddings for generated utterances using different perturbation methods. All the generated utterances are perturbed from the same source utterance. \textbf{Real} denotes speaker embedding of real utterances sampled from VCTK test set.}
  \label{fig:tsne}
\end{figure}

We argue that both the diversity and naturalness of generated utterances' timbre will affect the final performance.
To explain our method's effectiveness, we use t-SNE \cite{van2008visualizing} to visualize speaker embeddings of perturbed utterances in Fig. \ref{fig:tsne}.
All the perturbed utterances are based on the same source utterance.
The results reveal that the utterances generated by our method exhibit the greatest speaker diversity. Additionally, we plot the speaker embeddings of real utterances randomly selected from the test set. The distribution of speaker embeddings from our method closely resembles that of the real data.
In contrast, the timbre of utterances generated by the \textbf{NANSY} method appears abnormal, with its distribution significantly deviating from that of the real data.
Perturbed utterances of different methods can be found in our demo page too.

\subsubsection{Impact of teacher models}
In this section, we study the impact of different teacher models.
Results for models \textbf{b1/b2/b3} indicate that they yield similar performance levels.
In traditional semi-supervised learning tasks, pseudo labels are typically treated as training targets, and it is generally observed that higher-quality pseudo labels lead to improved results. However, in our approach, we utilize pseudo utterances as input to the model. Consequently, the minor differences among the teacher models have a limited effect on the final performance.

\subsection{Ablation study of Speaker Sampling}

\begin{table}[t]
\caption{Ablation study of Speaker Sampling. $\boldsymbol{\alpha}$ denotes the probability of speaker sampling.}
\begin{center}
\begin{tabular}{|c|c|c|c|}
\hline
{Models} & {$\alpha$} & {WER($\downarrow$)} & {SECS($\uparrow$)} \\
\hline
b3 & 0 & 6.7 & 0.780 \\
\hline
c1 & 0.1 & 6.0 & 0.773 \\
\hline
c2 & 0.01 & 6.1 & 0.778 \\
\hline
\end{tabular}
\label{tab:ablation2}
\end{center}
\end{table}

We explore the impact of \textit{Speaker Sampling} in Table \ref{tab:ablation2}.
Results show that our model with speaker sampling \textbf{c2} can get a better WER compared to the baseline model \textbf{b3}.
It can be attributed to the alleviation of mismatch between training and inference.
However, the results of model \textbf{c1} reveal that speaker similarity will degrade if the probability $\alpha$ is big.
We argue that aggressive speaker sampling may increase the difficulty of model training.

In conclusion, the objective results demonstrate that \textit{Pseudo Conversion} enhances speaker similarity while \textit{Speaker Sampling} contributes to improved intelligibility.

\section{Conclusions}
In this work, we first analyze the mismatch problem between training and inference in the one-shot voice conversion task.
Then we propose a novel training strategy combining \textit{Pseudo Conversion} and \textit{Speaker Sampling} to mitigate the mismatch problem.
Finally we verify the effectiveness of our proposed method, which outperforms previous methods.

It is important to note that our method employs a two-stage training process, as the teacher voice conversion model must be trained initially. In future work, we aim to explore the development of a one-stage training framework.

\bibliographystyle{IEEEtran}
\bibliography{mybib}

\end{document}